\documentclass[twocolumn,prl,aps,superscriptaddress,showpacs]{revtex4}
\usepackage{mathptmx}
\usepackage{subfigure}
\usepackage{dcolumn}
\usepackage{amsmath,amssymb}
\usepackage{bm}
\usepackage{color}
\usepackage{latexsym}
\usepackage[english]{babel}
\usepackage{times}
\usepackage{psfrag,graphicx}
\usepackage{epsf}
\usepackage{amsfonts}
\usepackage{natbib}
\usepackage{epstopdf}
\usepackage{appendix}
\usepackage[colorlinks=true,citecolor=blue,linkcolor=blue]{hyperref}
\usepackage{ulem}

\begin{document}


 %%%%%%%%%%%%%%%%%%%%%%%%%%%%%%%%%%%%%%%%%%%%%%%%%%%%%%
 % Title 8/ 15 words
 %%%%%%%%%%%%%%%%%%%%%%%%%%%%%%%%%%%%%%%%%%%%%%%%%%%%%%

\title{Tunable multiple layered Dirac cones in optical lattices}

\author{Z. Lan}
\affiliation{SUPA, Department of Physics, Heriot-Watt University, EH14 4AS, Edinburgh, United Kingdom}

\author{A. Celi}
\affiliation{ICFO - The Institute of Photonic Sciences Av. Carl Friedrich Gauss, num. 3, E-08860 Castelldefels (Barcelona), Spain}

\author{W. Lu}
\affiliation{SUPA, Department of Physics, Heriot-Watt University, EH14 4AS, Edinburgh, United Kingdom}

\author{P. \"Ohberg}
\affiliation{SUPA, Department of Physics, Heriot-Watt University, EH14 4AS, Edinburgh, United Kingdom}

\author{ M. Lewenstein}
\affiliation{ICFO - The Institute of Photonic Sciences Av. Carl Friedrich Gauss, num. 3, E-08860 Castelldefels (Barcelona), Spain}
\affiliation{ICREA-Instituci\'{o} Catalana de Recerca i Estudis Avancats, 08010 Barcelona, Spain }

\begin{abstract}

We show that multiple layered Dirac cones can emerge in the band structure of properly addressed multicomponent cold fermionic gases in optical lattices. The layered Dirac cones contain multiple copies of massless spin-$1/2$ Dirac fermions at the {\it same }location in momentum space, whose different Fermi velocity can be tuned at will. On-site microwave Raman  transitions can further be used to mix the different Dirac species, resulting in either splitting of or preserving the Dirac point (depending on the symmetry of the on-site term). The tunability of the multiple layered Dirac cones allows to simulate a number of fundamental phenomena in modern physics, such as neutrino oscillations and exotic particle dispersions with $E\sim p^N $ for arbitrary integer $N$.

\end{abstract}

\maketitle

%%%%%%%%%%%%%%%%%%%%%%%%%%%
%Introduction
%%%%%%%%%%%%%%%%%%%%%%%%%%%

{\it Introduction}. The concept of a quantum emulator was first introduced by Feynman as a way to avoid the difficulty of simulating quantum phenomena with classical computers \cite{feynman82}. The idea was to use one controllable system to simulate another, possibly computationally intractable system. Nowadays Feynman's intuition is being implemented in various setups and among them, cold gases of neutral atoms play a central role \cite{lewe07,bloch08,buluta09}. Systems of trapped ultracold atoms in optical lattices have proven to be a remarkable tool for simulating a vast range of condensed matter, and lately also high energy physics phenomena. Graphene and  topological insulators have recently attracted a significant attention due to their massless low-energy excitations, termed Dirac fermions, which govern the electron transport properties~\cite{graphene_rev, top_insulators_rev}. They are examples of quasi-relativistic dynamics in a non-relativistic environment.
In an effort to extend such behavior to quasi-particles with arbitrary spin, a quantum simulator of the so called Dirac-Weyl fermions with arbitrary large spin~\cite{weyl} has been proposed. This can be implemented by using  multicomponent cold fermionic atoms trapped in an optical lattice, and  by tuning the hopping matrices between these internal states according to $\mathfrak{su}$(2) algebra representations. This setup allows us to assign an arbitrary spin to the emergent low-energy excitations, whose band structures display multiple layered Dirac cones at four isolated points in the first Brillouin zone, termed Dirac points. Such band structure gives rise to exotic properties, e.g., a rich anomalous quantum Hall effect and Klein multi-refringent tunnelling. The shape of the cones, i.e., the different Fermi velocities of quasi-particles are completely fixed by the $\mathfrak{su}$(2) representation. In this Letter, we show how to relax the above constraint, providing a playground for simulations of high energy phenomena such as neutrino oscillations \cite{neutrino1} in Lorentz and CPT  breaking/no breaking scenarios \cite{CPT,splitting,neutrino2}, and modifications of dispersion relations \cite{MDR,dispersion}.

{\it The model}. For an optical square superlattice filled with multicomponent alkali fermions, the Hamiltonian is \cite{superlattice_scheme}
 \begin{eqnarray}
\label{hamiltonian}
H=H_t+H_o=-\sum_{\boldsymbol{r},\nu}\sum_{\tau\tau'}t_{\nu}[\mathbb{T}_{\nu}]_{\tau'\tau}c_{\boldsymbol{r}
+{\nu},\tau'}^{\dagger}c_{\boldsymbol{r}\tau}+\text{H.c.}\nonumber\\+\sum_{\boldsymbol{r}}\sum_{\tau\tau'}[\mathbb{O}]_{\tau'\tau}c_{\boldsymbol{r}\tau'}^{\dagger}c_{\boldsymbol{r}\tau},
\end{eqnarray}
where $H_t$ describes the nearest-neighbor hopping dynamics, $H_o$ the on-site dynamics,  and $c_{\boldsymbol{r}\tau}^{\dagger}(c_{\boldsymbol{r}\tau})$  are fermionic creation (annihilation) operators with an internal index $\tau$.  The optical potential is assumed to be deep enough so that free hopping due to kinetic energy is suppressed, and all system dynamics is induced by Raman transitions. In particular the hopping along the $\nu=x,y$ direction is described by a spin-dependent operator $\mathbb{T}_{\nu}$ while the on-site spin-flipping is described by $\mathbb{O}$.  For a detailed discussion of experimental realizations we refer the reader to Refs. \cite{weyl,superlattice_scheme}.  In such laser-assisted hopping schemes, each of the matrix elements of the hopping operators is realized via an effective four-photon process, where the spin of the atom can be flipped with at most $|\Delta m_F|=4$~\cite{superlattice_scheme}. As explained below, we demand non-zero elements of the hopping matrices only around the diagonal, i.e., with a superdiagonal and a subdiagonal which involve only $|\Delta m_F|=1$.

{\it Tunneling dynamics and multiple layered Dirac cones.} In momentum space $H_t$ reduces to $H=\sum_{\boldsymbol{k}}\Psi^{\dagger}(\boldsymbol{k})H(\boldsymbol{k})\Psi(\boldsymbol{k})$ where $H(\boldsymbol{k})=-\sum_{\nu}2t_{\nu}\mathbb{T}_{\nu}\cos(k_{\nu})$, the spinor $\Psi(\boldsymbol{k})=(c_{\boldsymbol{k}1},...,c_{\boldsymbol{k}n})^{T}$ contains the fermionic operators, and $k_\nu$ is in units of the lattice spacing.  When the hopping operators $\mathbb{T}_{\nu}$ are tuned according to the  $n$-dimensional representation of the $\mathfrak{su}(2)$ Lie algebra, namely,
$\mathbb{T}_x= \Sigma_x$ and $\mathbb{T}_y= \Sigma_y$ which fulfill the corresponding algebra $[\Sigma_z,\Sigma_{\pm}]=\pm 2\Sigma_{\pm}$ and $[\Sigma_x,\Sigma_y]=2i \Sigma_z$ where
$\Sigma_{\pm}=\Sigma_x\pm i \Sigma_y$, then the low energy excitations of the system are Dirac-Weyl fermions with integer or half-integer spin depending on whether $n$ is odd or even. A simple representation of $\Sigma_x$ and $\Sigma_y$ for spin $s$ with $n=2s+1$ which generalizes the Pauli representation for spin $1/2$, can be expressed in terms of a $(n-1)$-vector $\boldsymbol{\rho}_{n}\equiv(\rho_1, \rho_2, \cdots, \rho_{n-1})$ with $\rho_j=\sqrt{j(n-j)}$  such that $\Sigma_x=\{\textrm{superdiag}\{\rho_j\}, \textrm{subdiag}\{\rho_j\}\}$ and $\Sigma_y=\{\textrm{superdiag}\{-i\rho_j\}, \textrm{subdiag}\{i\rho_j\}\}$, $j=1,\dots,n-1$.\\
$\phantom{T}$It is worth noticing that for any real vector $\boldsymbol{\rho}$, $\mathbb{T}_x= \{\textrm{superdiag}\{\rho_j\}, \textrm{subdiag}\{\rho_j\}\}$ and $\mathbb{T}_y= \{\textrm{superdiag}\{-i\rho_j\}, \textrm{subdiag}\{i\rho_j\}\}$, the location of the Dirac points of $H(\boldsymbol{k})$ does not change because the spectrum is $E(\boldsymbol{k})=\epsilon_{\boldsymbol{\rho}}|\boldsymbol{g}_{\boldsymbol{k}}|$ where $\epsilon_{\boldsymbol{\rho}}$ are the eigenvalues of the $\mathbb{T}_\nu$ matrix \cite{ph symmetry},  and $\boldsymbol{g}_{\boldsymbol{k}}=(2t_x\textrm{cos}k_x, 2t_y\textrm{cos}k_y, 0)$. This observation is crucial as the {\it leitmotif} of the Letter is to preserve the stability of the Dirac points while relaxing the integer or half-integer spin structure such that we can construct arbitrary effective speeds of light in the emergent quasi-relativistic scenario. With this extension, properties that depend on the topological aspects of the system such as the topologically invariant Berry phase, the topological charge and the Hall plateaus will consequently not change, while properties depending on local aspects of the system such as the butterfly nature of the spectrum and the Klein multirefringent tunnelling  will change \cite{weyl}. The resulting spectrum at each Dirac point will be a collection of Dirac fermions with tunable effective speeds of light and a topological charge $N$ equal to the number of the layers, i.e. $N=s+1/2$ for half-integer spin.

As an example,  we consider a double layered Dirac cone structure with tunable Fermi velocity for each component. We parametrize $\boldsymbol{\rho}_4 $ as $\rho (\textrm{sin}\theta \textrm{cos}\varphi, \textrm{sin}\theta \textrm{sin}\varphi, \textrm{cos}\theta) $. The corresponding spectrum is of the form
$\epsilon=\rho\sqrt{1\pm\chi/2}/\sqrt{2}$, where $\chi=\sqrt{3+\cos 2\varphi+\cos 4\theta-\cos 2\varphi\cos 4\theta}$. Interestingly, a birefingent breakup of the doubly degenerate Dirac cones into cones with different speeds of light have also been discussed in quite a different setup with cold atoms ~\cite{birefringence}.  It is worth stressing that our setup allows a breakup, or indeed a coalition, of {\it any} number of Dirac fermions.
For example, when $\chi=0$ the two layered cones collapse into a single degenerate one with topological charge $N=2$. This is also the mechanism for mixing the different Dirac species when on-site dynamics is introduced. The $N=3$ situation, on the other hand, can be used to mimic the three families of fermions in particle physics. These hopping matrices considered above can also be seen as generalized spin-orbit coupled systems (see \cite{SOC} for a recent realization in a Bose-Einstein condensate).

\begin{figure}
\centering
	\includegraphics[width=0.9\columnwidth]{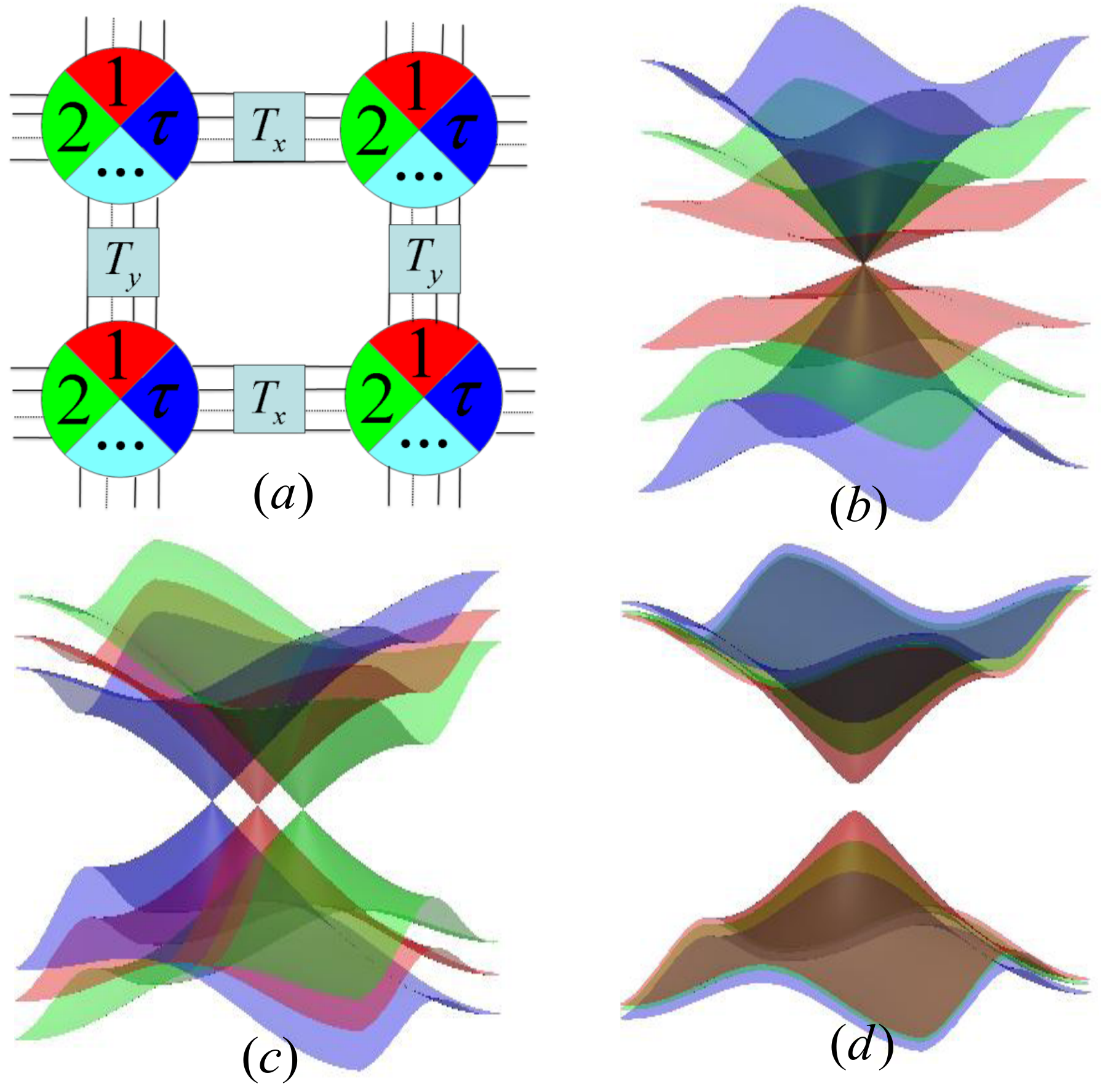}
	\caption{\label{figure1} Setup and typical band structures considered in the text. (a) Schematic illustration of the optical lattice and its tunneling mechanism. (b) The triple layered Dirac cones with three different effective speeds of light. (c) The splitting of triple degenerate Dirac cones. (d) Gap opening of triple degenerate Dirac cones. The band structures in (c) and (d) are used to simulate the exotic and traditional neutrino oscillations as discussed in the text.}
\end{figure}

  %%%%%%%%%%%%%

 {\it On-site dynamics and Topological Phase Transitions (TPT)}. While the hopping Hamiltonian $H_t$ in Eq. (\ref{hamiltonian}) allows us to create a collection of Dirac fermions at the {\it same} location in momentum space with tunable Fermi velocities, a constant on-site term $H_o$ gives an effective mixing of the different Dirac species. This mixing mechanism is fundamentally different compared to the mixing mechanism in graphene. In graphene the Dirac fermions are located at {\it different} sites $K_{\pm}$ in the momentum space. A commensurate perturbation $G=K_+-K_-$ introduced by lattice distortions is needed to mix the Dirac fermions (see~\cite{chiralGraphene1, chiralGraphene2} for a discussion of the chiral mixing in graphene by Kekule texture when constructing the chiral gauge theory of graphene, and \cite{unrealgraph,dora} for arbitrary spin quasi-particles). Furthermore, opening a gap in monolayer graphene is difficult. Our setup allows mixing of any number of Dirac fermions of the same chirality, and to open up gaps simply by the on-site microwave Raman transitions.

We now consider the manipulation of  triple degenerate Dirac cones with a topological charge of $N=3$ at a single Dirac point $\boldsymbol{K}=\frac{\pi}{2}(1,1)$, i.e., at the same location in momentum space. We write the constant on-site term - a special case  of the $\mathbb{O}$ of Eq. (\ref{hamiltonian}) - in the form of $\boldsymbol{h}^i\cdot  \boldsymbol{\sigma}$ where $\boldsymbol{h}^i=(h_x^i,h_y^i,h_z^i)$, $i=1,2,3$, are constant vectors, and $\boldsymbol{\sigma}=(\sigma_x,\sigma_y,\sigma_z)$ is a vector of Pauli matrices. The unperturbed case $\boldsymbol{h}^i=0$ corresponds to $\boldsymbol{\rho}_{6}=(1,0,1,0,1)$. The result is a combination of mixing and splitting of the Dirac points, which are the paradigm of topological quantum phase transitions. The corresponding Hamiltonian near the Dirac point is
\begin{equation}
 \label{N=3}
 H_m (\boldsymbol{k})=\left[\begin{array}{ccc} (\boldsymbol{g}_{\boldsymbol{k}}-\boldsymbol{h}^1) \cdot\boldsymbol{\sigma} & 0 &0 \\
 0& (\boldsymbol{g}_{\boldsymbol{k}}-\boldsymbol{h}^2) \cdot\boldsymbol{\sigma} &0 \\
0&0&  (\boldsymbol{g}_{\boldsymbol{k}}-\boldsymbol{h}^3) \cdot\boldsymbol{\sigma}
\end{array}\right].
\end{equation}

The spectrum is given by $E_i=\pm|\boldsymbol{g}_{\boldsymbol{k}}-\boldsymbol{h}^i|$. This expression provides a convenient way to manipulate each of the three Dirac species. While $h^i_x$ and $h^i_y$ control the positions of the split Dirac points, $h^i_z$ controls the gap of the energy spectrum (see Fig.~\ref{figure1}). By requiring $E=0$ and setting $h^i_z=0$ for the time being, the split Dirac points are determined by the conditions  $2t_x\textrm{cos}k_x- h^i_x=0$ and $2t_y\textrm{cos}k_y- h^i_y=0$. In principle, $\boldsymbol{h}^i$ may be tuned to appropriately large values to create marginal Dirac points with topological charge $N=0$ by merging the Dirac points of opposite chirality. The applications of such a scenario go beyond the realm of condensed matter \cite{helium}. However, in what follows we will focus on perturbative splitting.

\begin{figure}
\centering
	\includegraphics[width=0.9\columnwidth]{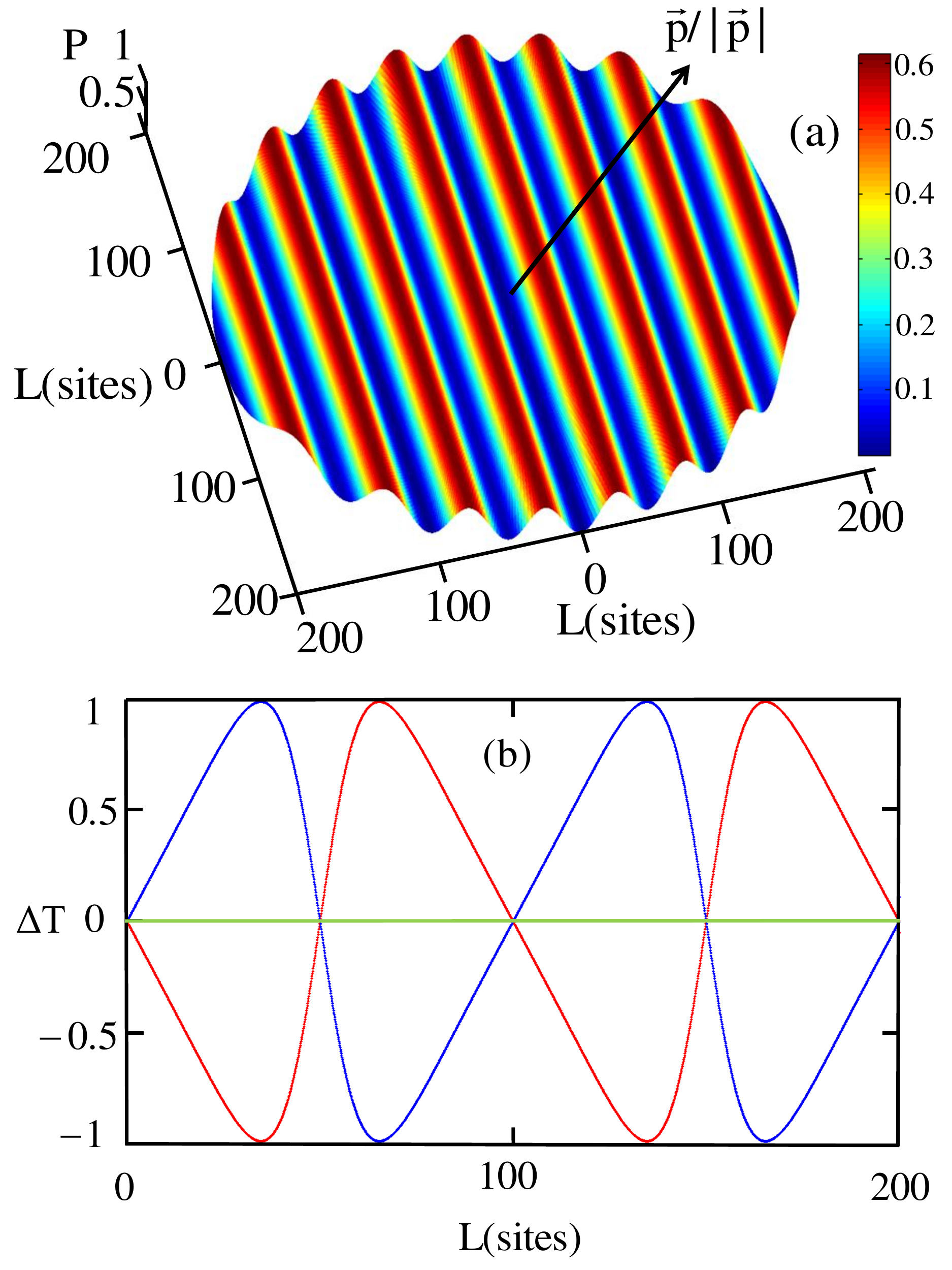}
	\caption{\label{figure2}  Anisotropic quasi-neutrino oscillations and T-violations in optical lattice with mixing angles $(\theta_{12},\theta_{13},\theta_{23})=(1/4,1/4,1/4) \pi$ and splitting $\boldsymbol{h}=0.01(2\pi)(1,0,0) $. (a) The oscillations  of the probability $P(\nu_{e} \rightarrow \nu_{\mu})$ with the directional vector $\hat{p}$ of the momentum where  the CP-violation phase is  $\delta=0$ . (b) The T-violation behaviour with $\delta=\pi/2$ (red), $3\pi/2$ (blue), $0, \pi$ (green), where $\Delta T=(P(\nu_{e} \rightarrow \nu_{\mu})-P(\nu_{\mu} \rightarrow \nu_{e}))/(P(\nu_{e} \rightarrow \nu_{\mu})+P(\nu_{\mu} \rightarrow \nu_{e}))$.}
\end{figure}

{\it Neutrino oscillations.} Neutrino oscillations (NO) are considered by many as a possible window for physics beyond the Standard Model. The most accepted mechanism to explain such oscillations is in terms of the nontrivial mass term matrix. The Lagrangian density describing the flavour states $\boldsymbol{\nu}^T_f=(\nu_e, \nu_{\mu},\nu_{\tau})$ with a mixing mass term $M$ is $\mathcal{L}=\overline{\boldsymbol{\nu}}_f(x)(i\gamma^{\mu}\partial_{\mu}-M)\boldsymbol{\nu}_f(x)$.  The flavour of the neutrino oscillates as it propagates since the flavour eigenstates do not coincide with their mass eigenstates with a definite mass and energy.
In general,  the weak interaction flavour eigenstates can be represented as a coherent linear superposition of the mass eigenstates, $\boldsymbol{\nu}_f=U\boldsymbol{\nu}_m$ where  $\boldsymbol{\nu}^T_m=(\nu_1, \nu_2,\nu_3)$ describes the state with definite masses and $U$ is the unitary transform matrix that diagonalizes the mixing mass matrix $M$, known as the Pontecorvo-Maki-Nakagawa-Sakata (PMNS) lepton mixing matrix. For all currently observed NO, the corresponding masses are less than 1$eV$ and the energies are at least $E\gtrsim 1 MeV$, with a Lorentz factor greater than $10^6$ in all cases. In this ultrarelativistic limit the energy is given by $E_i=(p^2c^2+m_i^2c^4)^{1/2}\simeq E+m_i^2c^4/2E$, $E=c|p|$. The time-dependent mass states are $|\nu_i(t)\rangle=\exp[-iE_it/\hbar]|\nu_i\rangle$, thus $ \boldsymbol{\nu}_f(t)=UTU^{\dagger}\boldsymbol{\nu}_f(0)$ where $T=\textrm{diag}\{\exp[-iE_1t/\hbar],\exp[-iE_2t/\hbar],\exp[-iE_3t/\hbar] \}$ is the time evolution matrix, i.e., $P(\nu_{\alpha} \rightarrow \nu_{\beta})= |\sum_j U_{\alpha j}U_{\beta j}^{\ast} \exp[-i\frac{m_j^2c^4}{2\hbar E}t]|^2$,  which shows that the neutrino flavour changes with time due to the mass difference between the mass states ~\cite{neutrino1}. The question of which mechanism generates such mass is still not completely settled.

{\it NO as TPT in optical lattices.} An analogue of NO can be engineered in optical lattices (for a classical analogue of neutral-meson oscillation see \cite{kostelecky00}).
The simple model of Eq. \ref{N=3} is the starting point to capture the essence of  3 species' mixing. Interestingly, our setup allows us to simulate both traditional NO and exotic anisotropic NO. In the following, we will consider an ultra-relativistic limit for coupled pseudo-particles, i.e.,  $|\boldsymbol{h}|\ll|\boldsymbol{g}_{\boldsymbol{k}}|\simeq c_l |\boldsymbol{p}|$, where  $\boldsymbol{p} \equiv (\boldsymbol{K} -\boldsymbol{k})$.  For the traditional NO, the analogue of a mass term can be reproduced by setting $\boldsymbol{h}^i=(0,0,h_z^i)$ for which the spectrum $E_i=|\boldsymbol{g}_{\boldsymbol{k}}-\boldsymbol{h}^i|$ reduces to $E_i\simeq c_l|\boldsymbol{p}| + {(h_z^i)}^2/(2 c_l |\boldsymbol{p}|)$, thus $h_z^i$ represents the mass of the neutrinos. Moreover, our setup also allows us to simulate exotic anisotropic NO where a {\it direction dependent mass} can be achieved in the limit of small $\boldsymbol{h}^i=(h_x^i,h_y^i,0)$ which we shall focus on in this study. In this case, the spectrum of $E_i=|\boldsymbol{g}_{\boldsymbol{k}}-\boldsymbol{h}^i|$ reduces to $E_i\simeq c_l(|\boldsymbol{p}|+\hat p \cdot \boldsymbol{h}^i)$, where $\hat p \equiv \boldsymbol{p}/(c_l|\boldsymbol{p}|)$. For sake of simplicity, we consider $\boldsymbol{h}^1=0$ and  $\boldsymbol{h}^2=\boldsymbol{h}=-\boldsymbol{h}^3$.
The $H_m (\boldsymbol{k})$ in Eq. \ref{N=3} therefore plays the role of the block diagonal Hamiltonian in the {\it mass} basis. In fact, to observe NO in the lab we have to implement the Hamiltonian $H_f(\boldsymbol{k})$ governing the {\it flavour} pseudo-particles. In terms of the PMNS mixing matrix $U$, $H_f (\boldsymbol{k})=(U^\dagger\otimes I_2) H_m (\boldsymbol{k}) (U\otimes I_2)  = I_3\otimes \boldsymbol{g}_{\boldsymbol{k}}\cdot \boldsymbol{\sigma}+ M \otimes\boldsymbol{h}\cdot \boldsymbol{\sigma}$, where $I_n$ is the $n\times n$-identity matrix, and $M\equiv U^\dagger \textrm{diag}\{0,-1,1\} U$, i.e., $M_{ij}=U_{3j}U_{3i}^{*}-U_{2j}U_{2i}^{*}$.  In Lagrangian terms, our 2+1 model is $\mathcal{L}=\overline \Psi_f(x)(i\gamma^{\mu}\partial_{\mu}-M \gamma^\mu h_\mu )\Psi_f(x)$ near the Dirac point. Thus the flavour Hamiltonian  fits in the family of Eq. \ref{hamiltonian}, $H_f (\boldsymbol{k})=H_t(\boldsymbol{k})+H_o(\boldsymbol{h})$, with $H_t(\boldsymbol{k})=I_3\otimes \boldsymbol{g}_{\boldsymbol{k}}\cdot \boldsymbol{\sigma} $ and $H_o(\boldsymbol{h})= M \otimes\boldsymbol{h}\cdot \boldsymbol{\sigma}$. Remarkably, while $H_t(\boldsymbol{k})$ can be implemented by the triple-layered degenerate Dirac cones as discussed above, the PMNS mixing matrix is completely encoded by the on-site Hamiltonian $H_o(\boldsymbol{h})$, which allows us in principle to simulate any mixing angles and CP-violating phase - the standard parametrization of the PMNS can be found for instance in \cite{neutrino1} - by the on-site microwave Raman transitions \cite{superlattice_scheme}.
As an explicit example, we discuss the simplified case when all the mixing angles are equal to $\pi/4$ with
\begin{equation}
 \label{Hfig}
 H_f^{\text{ex}}=\left[\begin{array}{ccc} (\boldsymbol{g}_{\boldsymbol{k}}- \cos\frac{\delta}{\sqrt 2}\boldsymbol{h})\cdot\boldsymbol{\sigma} & -i \sin\frac{\delta}{\sqrt 2}\boldsymbol{h}\cdot\boldsymbol{\sigma} &\frac 12 \boldsymbol{h}\cdot\boldsymbol{\sigma}\\
 i \sin\frac{\delta}{\sqrt 2}\boldsymbol{h}\cdot\boldsymbol{\sigma}&(\boldsymbol{g}_{\boldsymbol{k}}+ \cos\frac{\delta}{\sqrt 2}\boldsymbol{h})\cdot\boldsymbol{\sigma} &-\frac 12 \boldsymbol{h}\cdot\boldsymbol{\sigma} \\
\frac 12 \boldsymbol{h}\cdot\boldsymbol{\sigma}&-\frac 12 \boldsymbol{h}\cdot\boldsymbol{\sigma}&  \boldsymbol{g}_{\boldsymbol{k}}\cdot\boldsymbol{\sigma}
\end{array}\right],
\end{equation}
and $\delta$ is the CP-violating phase.
 The quasi-neutrino oscillations are shown in Fig. \ref{figure2}. In  Fig.~\ref{figure2} (a), the oscillation probabilities for $\delta=0$ are plotted against $\hat{p}$, showing anisotropic and energy-independent behaviors. 
 
 The Hamiltonian of Eq. \ref{Hfig} is real when $\delta=0$, hence invariant under time-reverse symmetry $T$. Since the oscillation period is inversely proportional to the splitting,  one obtains for a small splitting parameter $|\boldsymbol{h}| \sim 0.01 (2\pi/a)$, a period which is of the size of typical lattices. In high energy physics, evidence for NO is presented over a distance $L\geq 100km$ and no evidence for oscillations for $L\leq 1km$ \cite{neutrino1}.  In Fig.~\ref{figure2} (b), the effect of $\delta\neq0$ is considered. The Hamiltonian becomes complex and $T$ is violated. Due to CPT invariance, T-violation is equivalent to CP-violation. The Time-violation behaviour of the transition probability $P(\Psi_{\alpha} \rightarrow \Psi_{\beta})\neq P(\Psi_{\beta} \rightarrow \Psi_{\alpha})$ is shown for $\delta=\{0,1/2,1,3/2\}\pi$. The signature of CP-violation is rather spectacular.
 In the optical lattice experiment, direct evidences of the above phenomena can be obtained by measuring the different populations in different points of the lattice by the colour resolution or individual atom detection techniques \cite{SAD1,SAD2}. Lorentz and CPT conservation hold if $\boldsymbol{h}$ is covariant.

{\it Lorentz and CPT violation in optical lattices.} Different sources of Lorentz and CPT violating oscillation terms can be considered in optical lattices. The easiest is to simulate anisotropic Minkowski space-time, i.e., $t_x \neq t_y$, thus $c_x \neq c_y$. 
% Anisotropic and isotropic 
Alternatively, any isotropic and anisotropic Lorentz violating mass term \cite{CPT}, $\mathcal{M}=m+im_5\gamma_5+a^{\mu}\gamma_{\mu}+b^{\mu}\gamma_5\gamma_{\mu}+H^{\mu\nu}\delta_{\mu\nu}$, can be conveniently engineered. 
%iso
In the former case $m$ and $m_5$ are Lorentz violating if they have an energy dependence \cite{isoCPT}, and their CPT behavior depends on their origin.  Phenomenological models including terms such as $m\propto E^p$, with $p>0$, are very appealing, and induce strong modifications of the oscillations at high energy. In optical lattices, such terms can be induced by time-dependent Bragg scattering. A detail study is left for a future work.   
%aniso
 In the latter case, for constant coupling,  
$H$ is CPT conserving but Lorentz violating whereas the $a$ and $b$ terms are both CPT and Lorentz violating. In principle all of them can be reproduced by the more complicated on-site spin-flipping Hamiltonian as in Eq. \ref{hamiltonian}. However, the role of symmetries in the analogue simulation is subtle as they are {\it artificial}. The Hamiltonian engineered in the lattice corresponds to a certain reference frame (or gauge) only. To consider transformed Hamiltonians in other frames implies physical modifications of the experimental apparatus - for similar discussions of gauge symmetries see \cite{Boada09}. In practise, we can choose how $\mathcal{M}$ transforms. 
%sensibility
The present bounds on Lorentz violation \cite{CPTtable} in the accessible energy regime are very tight, and far from the current accuracy achievable in optical lattices - the major limitation is due to the lattice correction to the continuum theory, hence $\sim 1/(\# $ lattice points$)$. However, the simulator is a useful tool as the parameters can be tuned almost at will and dynamically generated flavour couplings and further strong coupling interactions \cite{thirring} may be included.

{\it Helicity projection and mixing with anti-neutrinos.} Above, we have considered a single Dirac point $\boldsymbol{K}=\frac{\pi}{2}(1,1)$. Since the system involves both positive and negative helicities \cite{weyl}, we briefly discuss how to achieve helicity projection for completeness. We note this issue is closely related to the so called {\it valley polarization} ~\cite{valleytronics1} in the graphene community where the degeneracy of the two inequivalent Dirac points related to time-reversal symmetry is lifted such that only a single valley is occupied. A specially designed optical potential barrier may be used to filter out one valley following the discussion in \cite{valleytronics2,valleytronics3}, thus achieving the helicity projection. Alternative method would be to engineer a Wilson mass term in the optical lattices which decouple doublers thus leaving a single Dirac point \cite{Wilson_mass}. In order to consider the mixing between neutrino and anti-neutrino, a Bragg pulse that connects the positive and negative helicities may be used as described in the $(3+1)$ Dirac fermions in optical lattices ~\cite{Bragg_pulse}. An alternative approach is to couple a two layered system where the sign of the topological charge in one layer is flipped, by sending $\mathbb{T}_{\nu}\to-\mathbb{T}_{\nu}$.

{\it Engineering exotic particle dispersions.} Modifications of dispersion relations have attracted a lot of interest in both condensed matter and high energy physics. The Dirac fermions in graphene and semi-Dirac fermions with linear dispersion in one direction and quadratic dispersion \cite{semiDirac} in the other are examples of engineered dispersion relations in a condensed matter setting.  On the other hand, in high energy physics, modifications of the energy momentum dispersion relations at the Planck scale are suitable, for instance, to reconcile Lorentz symmetry and finite resolution of the spacetime points in quantum gravity models \cite{MDR}. As an alternative to the splitting of the multiple degenerate Dirac point, it is possible to preserve the Dirac point and to engineer exotic particle dispersion to have any power of momentum. This Dirac-point preserving mixing is described by the Hamiltonian $ \mathcal{H}(\boldsymbol{p})=\{ \textrm{superdiag}\{g \sigma^+\},\textrm{diag} \{\boldsymbol{p}\cdot\boldsymbol{\sigma} \}, \textrm{subdiag}\{g \sigma^-\} \}$, which mixes the $N$ Dirac species and  gives the spectrum $E\sim p^N$ ~\cite{dispersion}. Our setup also allows us to investigate the mixing of Dirac-Weyl fermions with high spin beyond the above spin $1/2$ case, for instance the fate of the flat band with integer spin $s$ after mixing. Furthermore, disorder can also be introduced to the optical lattice~\cite{disorder} to simulate spacetime fluctuations at Planck scale, thus giving an opportunity to resolve possible conundrums in quantum gravity models by a combination of concepts, such as modifications of dispersion relations,  spacetime fluctuations, and Lorentz violations discussed above.

 {\it Conclusions.}  We have shown how multiple-layered Dirac cones can be used to simulate various phenomena from the realm of particle physics. The models discussed here deal with a quadratic Hamiltonian, but quartic \cite{thirring} and dynamical gauge \cite{dymgauge} interactions may be included. This, together with the possibility of changing the physical parameters in a wide range due to the advances in manipulation of cold atoms in  optical lattices, makes the quantum simulator a useful tool to test fascinating ideas from condensed matter to high energy physics. Furthermore, such quantum simulations can provide a guide to design nano devices, based on novel physical effects.
  We believe the interplay between condensed matter, particle physics and quantum simulations could lead us to new paradigms, much in the spirit of fundamental theories as emerging phenomena \cite{Wen}.

\acknowledgements {Z.L. acknowledges the support from SUPA. A.C. and M.L. acknowledge funding from
the Spanish MEC projects TOQATA (FIS2008-00784),
QOIT (Consolider Ingenio 2010), ERC Advanced Grant
QUAGATUA, EU STREP NAMEQUAM. M.L. acknowledges the financial support of Alexander von
Humboldt Foundation. P.\"O acknowledges support from the Carnegie Trust for the Universities of Scotland.}

{\it Note added.} Once this work was completed, \cite{NOiontrap} appeared proposing an ion trap simulator of NO in 1+1 dimensions.


\begin{references}
\bibitem{ph symmetry} For $\mathbb{T}_x=\{\textrm{superdiag}\{\rho_j\}, \textrm{subdiag}\{\rho_j\}\}$, the spectrum is determined by the zeros of  $det(\mathbb{T}_x-\epsilon \mathbb{I}^{(n)})=P_n(\epsilon)$ where $\epsilon$ is the eigenvalue. Consequently we get the recurrence relation $P_{n+1}(\epsilon)=-\epsilon P_n(\epsilon)-\rho_n^2 P_{n-1}(\epsilon) $,
 with the initial conditions $P_0=1$ and $P_1=-\epsilon$.  By induction we get $P_n(\epsilon)  =  f_n(\epsilon^2), n \in \{2k, \forall k\in \mathbb{Z}_+\}$ and
$P_n(\epsilon)  =   \epsilon g_n(\epsilon^2), n \in \{2k+1, \forall k\in \mathbb{Z}_+\}$, which demonstrates the particle-hole symmetry of $\mathbb{T}_x$, and the zero energy flat band for odd dimensional representations. It is straightforward to show that $\mathbb{T}_y$ has the same eigenvalues as $\mathbb{T}_x$ as they share the same characteristic polynomial.
\end{references}
\end{document}